\documentclass[11pt,a4paper,oneside]{article}

\usepackage[latin1]{inputenc}    
\usepackage{ngerman}             
\usepackage[english]{babel}      
\usepackage{amsmath}
\usepackage{amssymb}
\usepackage{amsfonts}
\usepackage{amsthm}
\usepackage{mathrsfs}
\usepackage{empheq}
\usepackage{exscale}             
\usepackage[final]{graphicx}     
\usepackage{flafter}
\usepackage{pict2e}
\usepackage[format=default,font=footnotesize,singlelinecheck=false,labelfont=bf]{caption}

\allowdisplaybreaks              

\renewcommand{\vec}[1]{\boldsymbol{#1}}

\DeclareMathOperator{\dd}{d\!}

\newcommand{\gap}{\medspace\negthinspace}
\newcommand{\abs}[1]{\left|#1\right|}
\newcommand{\totald}[2]{\frac{\dd #1}{\dd #2 }}       

\begin{document}

	\begin{center}
	\textbf{\huge On force-field models of the spacecraft flyby anomaly}\\[1cm]
		Wolfgang Hasse\footnote{Technical University Berlin, Institute for Theoretical Physics, Sekr. EW 7-1, Hardenbergstraße 36, 10623 Berlin, Germany}$^{,}$\footnote{Wilhelm Foerster Observatory Berlin, Astrometry Research Group, Munsterdamm 90, 12169 Berlin, Germany}, 
		Emrah Birsin$^{2,}$\footnote{Humboldt University of Berlin, Department of Physics, Newtonstraße 15, 12489 Berlin, Germany}, 
		Philipp Hähnel$^{2,3}$\\[.5cm]
		\today
	\end{center}

\section*{Abstract}

Recently, Anderson et al. \cite{anderson} published an empirical prediction formula for the so far unexplained parts of the velocity changes of spacecrafts during Earth flybys. In the framework of a perturbational approach, we show that there is no velocity-independent force field of the Earth - in addition to its Newtonian gravity field - that is to reproduce this formula. However, we give examples for fields modeling exactly the flyby anomaly which are quadratic functions of the velocity of the spacecraft.

\section{Introduction}

The flyby anomaly is an unexplained frequency change in the radio Doppler data during an Earth flyby of a spacecraft. Interpreted as a ``real'' effect, this corresponds to an anomalous velocity change by a few mm/s as we can see, for example, in Table 2 in the paper of Lämmerzahl, Preuss and Dittus \cite{lamm}, which gives an overview of this phenomenon.

Recently, Anderson et al. \cite{anderson} published an empirical prediction formula (eq. (2) in \cite{anderson}, eq. (1) in section 2 of the present paper), which accurately reproduces the measured anomalous velocity changes for all of the up to now six flybys for which sufficient data are available. Despite the fact that this formula is not related to any explanation of the phenomenon, it is a valuable progress for two reasons. Firstly, it makes sure that the anomalous velocity changes are not only significant, but there is a clear regularity in the data that rules out any exclusively statistically disturbing influences on the spacecraft orbits. Secondly, the special form of the discovered regularity enables us to systematically draw conclusions about the form of the unknown forces. Especially, if we assume that there is some unknown force field generated by the Earth (in addition to its Newtonian gravity field) and acting on the spacecrafts, we are no longer in the situation to guess the form of this field and then check whether it explains the measured anomalous velocity changes in all six cases. Instead, we can try to find force fields reproducing this formula by much more efficient mathematical methods. Certainly, this has to be done under the assumption that the prediction formula applies to all possible Earth flybys (at least in a wide range of the relevant orbital parameters). Finally, we can draw conclusions about the general properties of all such fields.

In the following section the latter will be done, resulting in strong restrictions for those fields. In particular, we prove that any force which depends only on the spacecraft's position in space and not on the spacecraft's velocity (like the Newtonian gravity) is strictly ruled out. On the other hand, section 3 demonstrates there is a wide variety of force fields being quadratic functions of the spacecraft's velocity and modelling exactly the flyby anomaly. In section 4 we discuss our results in view of the consequences for the possible nature of the unknown kind of interaction and for the energy transfer process.

\section{Exclusion of velocity-independent force fields}

The approximation formula for the anomalous part $\Delta V_\infty$ of the velocity change given by Anderson et al. \cite{anderson} reads
	\begin{align}
		\label{eq:caltech}
		\frac{\Delta V_\infty}{V_\infty} = \frac{1}{2}\frac{\Delta E}{E} = K(\cos\delta_i-\cos\delta_o)\,,
	\end{align}
where $V_\infty$ denotes the hyperbolic excess velocity of the spacecraft; $E$ and $\Delta E$, the total specific energy and the anomalous part of its change respectively, and $\delta_i$ and $\delta_o$, the declinations of the incoming and outgoing asymptotic velocity vectors. (The declinations, as well as $V_\infty$ and $E$, are defined as osculating Kepler orbital elements.) $K$ is a dimensionless constant which is in good agreement with twice the special relativistic $\beta$-factor of the Earth's spin rotation at the equator:
	\begin{align}
		\label{eq:K}
		K & = 2\frac{\omega_E R_E}{c}\approx3.099\cdot 10^{-6}
	\end{align}
($\omega_E$, $R_E$: angular rotational velocity and mean radius of the Earth. Even if we regard values specific for the Earth, hereafter we will not subscript the Earth's mass and thus it is $M_E\equiv M$.)

Our goal is to answer the following question: What is the mathematical form of a (classical,) stationary force field acting on a spacecraft (additionally to the Newtonian gravity field) that reproduces formula \eqref{eq:caltech}? Since the anomalous accelerations necessary to yield the measured anomalous velocity changes are only of the order $10^{-5}$ of the Newtonian gravitational acceleration \cite{lamm, lamm2}, it seems to be sufficient to treat the additional forces as small perturbations along the classical Kepler trajectories of the spacecrafts. In this section, we will restrict our considerations to forces that only depend on the spacecraft's position in space but not on the spacecraft's velocity. According to \eqref{eq:caltech}, such a field $\vec{F}$ must yield a total specific energy change
	\begin{align}
		\label{eq:energy}
		\Delta E & = \int\limits_{-\infty}^{\infty}\vec{F}(\vec{r}(t))\cdot\vec{V}(t)\dd t = 2KE(\cos\delta_i - \cos\delta_o)
	\end{align}
for any hyperbolian Kepler orbit. The integration is to be taken along the unperturbed trajectory $\vec{r}(t)$ with velocity vector $\vec{V}=\dot{\vec{r}}$. (Here and in the following $E$, $\Delta E$, $\vec{F}$ etc. denote the specific quantities, what means that, as in \cite{anderson}, these values are divided by the spacecraft's mass. Therefore, i. e., in accordance with the structure of \eqref{eq:caltech} and \eqref{eq:energy}, the force has been replaced by the acceleration.)

The most general expression for the field $\vec{F} = \vec F(r,\alpha,\delta)$ with $r = \abs{\vec r}$ is
	\begin{align}
		\label{eq:force}
		\vec{F}(r,\delta,\alpha) = f_1(r,\delta,\alpha)\vec e_r + f_2(r,\delta ,\alpha)\vec e_\delta + f_3(r,\delta,\alpha)\vec e_\alpha\,,
	\end{align}
where $\vec e_r$, $\vec e_\delta$ and $\vec e_\alpha$ are the unit vectors belonging to the radial and the usual equatorial geocentric coordinate directions respectively (they form an orthonormal basis at each point of the three-dimensional space, and, of course, $\alpha$ is the right ascension.) In view of \eqref{eq:caltech} and \eqref{eq:K}, the source of the additional field should be in a today unknown relation to the rotation of the Earth. Concerning the question whether there is a solution to equation \eqref{eq:energy} at all, we will show now that without loss of generality we may assume that $f_2 = f_3 = 0$ and $f_1(r,\delta,\alpha) = \frac{f(\delta)}{r^2}$ with some symmetric function $f$, appealing to the symmetry properties of formula \eqref{eq:caltech}. (Note, however, that to any solution $\vec{F}$ there are more solutions without the symmetry properties of \eqref{eq:caltech}. One possibility to find such solutions is to add the gradient of a potential which does not share the symmetries and converges to a constant direction-independent value in the limit $r\rightarrow \infty$.)
\newline

	\underline{Step 1:}\newline
Due to the rotational symmetry of \eqref{eq:caltech}, with respect to the Earth's axis, for any solution $\vec{F}$ there are new solutions, obtained by shifting the $\alpha$-coordinate by an arbitrary constant value. By means of superposition of ``shifted'' solutions of the equation \eqref{eq:energy} (which is linear in $\vec{F}$), we can completely eliminate the $\alpha$-dependence of the functions $f_1$,$f_2$ and $f_3$ (within the limit of the superposition of an infinite number of solutions).
\newline

	\underline{Step 2:}\newline
Because of the invariance of \eqref{eq:caltech} under the transformation $\alpha\rightarrow -\alpha$ (transforming $\vec e_\alpha\cdot\vec{V}$ to $-\vec e_\alpha\cdot\vec{V}$), we know that for any solution $\vec{F}(r,\delta) = f_1(r,\delta)\vec e_r + f_2(r,\delta)\vec e_\delta + f_3(r,\delta)\vec e_\alpha$ the field $f_1(r,\delta)\vec e_r + f_2(r,\delta)\vec e_\delta - f_3(r,\delta)\vec e_\alpha$ is also a solution. Half the sum of both solutions has no $\vec e_\alpha$ component; so without loss of generality we may henceforth set $f_3\equiv 0$.
\newline

	\underline{Step 3:}\newline
Similarly, the mirror symmetry of \eqref{eq:caltech} with respect to the Earth's equatorial plane implies that with any solution $\vec{F}(r,\delta) = f_1(r,\delta)\vec e_r + f_2(r,\delta)\vec e_\delta$ the field $f_1(r,-\delta)\vec e_r - f_2(r,-\delta)\vec e_\delta$ is also a solution. Half the sum of both solutions has a symmetrical $\vec e_r$ part and an antisymmetrical $\vec e_\delta$ part; so without loss of generality we may assume that $f_1(r,\delta)$ is symmetric in $\delta$ whereas $f_2(r,\delta)$ is antisymmetric in $\delta$.
\newline

	\underline{Step 4:}\newline
Formula \eqref{eq:caltech} reveals another symmetry property, namely a scaling symmetry: Let us be given an arbitrary hyperbolic orbit parametrized by the polar angle $\varphi$ of the polar coordinates in the orbital plane, i. e., $\vec{r}(\varphi)$ with modulus
	\begin{align}
		\label{eq:rparam}
		r(\varphi) = \frac{p}{1+\varepsilon\cos\varphi}\,,
	\end{align}
where $p$ and $\varepsilon$ are the semi-latus rectum and the numerical exentricity, respectively. If we enlarge this orbit by a ``blow-up'' factor $b\in(0,\infty)$, we get a new orbit $b\vec{r}(\varphi)$ with total energy $\frac{E}{b}$ instead of $E$. The other terms on the right-hand side of equation \eqref{eq:energy} remain unchanged under this transformation; so the anomalous energy change $\Delta E$ transforms also with the factor $\frac{1}{b}$, which yields the scaling property
	\begin{align}
		b\int\vec{F}(b\vec{r})\cdot\dd\vec{r} = \frac{1}{b}\int\vec{F}(\vec{r})\cdot\dd\vec{r}
	\end{align}
with the factor b on the left-hand side coming from the streching of the integration path (both integrals are understood to go along the unenlarged $(b=1)$ path in space). Since this scaling property holds true for all hyperbolic Kepler orbits, the effective part of the field $\vec{F}$ contains its $r$ dependence only in form of a factor $\frac{1}{r^2}$, whereas other powers of $r$ do not contribute to the net energy change $\Delta E$. (Formally, one can work this out by a Laurent development of $\vec{F}$ with respect to $r$, followed by a comparison of the powers of $b$.)

Consequently, if there is any solution of equation \eqref{eq:energy}, we can get a new solution by discarding all parts but the $\frac{1}{r^2}$ part (irrespective whether the original solution is of the ``long'' form \eqref{eq:force} or of a reduced form after the steps 1, 2 or 3). Hence, going on from step 3, without loss of generality we assume that $\vec{F}$ is of the form
	\begin{align}
		\label{eq:force2}
		\vec{F}(r,\delta) = \frac{f_1(\delta)}{r^2}\vec e_r + \frac{f_2(\delta)}{r^2}\vec e_\delta\,,
	\end{align}
where $f_1$ is a symmetric and $f_2$ an antisymmetric function.
\newline

	\underline{Step 5:}\newline
Finally, we may even omit the $\vec e_\delta$ part of \eqref{eq:force2}, as can be seen by the following ``radialization'' procedure: Replace $\vec{F}(r,\delta)$ by $\vec{F}(r,\delta) + \nabla\frac{\Phi(\delta)}{r}$, with $\Phi$ being an antiderivative of $-f_2$. (Because of $\lim\limits_{r\rightarrow\infty}\frac{\Phi(\delta)}{r} = 0$, the additional term does not contribute to $\Delta E$.) In our spherical coordinates, the gradient term reads $\nabla\frac{\Phi(\delta)}{r} = -\frac{\Phi(\delta)}{r^2}\vec e_r + \frac{\Phi'(\delta)}{r^2}\vec e_\delta$ and further, together with \eqref{eq:force2} and $\Phi'=-f_2$, it gives $\vec{F}(r,\delta) + \nabla\frac{\Phi(\delta)}{r} = \frac{f_1(\delta)-\Phi(\delta)}{r^2}\vec e_r$. Thus, if there is any solution of the form \eqref{eq:force2}, there will be a solution of the form
	\begin{align}
		\label{eq:forcefinal}
		\vec{F}(r,\delta) = \frac{f(\delta)}{r^2}\vec e_r\,,
	\end{align}
too (with $f = f_1-\Phi$). Since $f_2$ is assumed to be antisymmetric, $\Phi$ is a symmetric function, which guarantees that the symmetry of the $\vec e_r$ part in $\delta$ is preserved under this manipulation. (Alternatively, by an analogous procedure, one can transform the $\vec e_r$ part of \eqref{eq:force2} to zero, leaving a purely tangential field.)
\newline

Now, to prove the velocity independence of any force field wrong, we assume there is a velocity-independent force field of the form \eqref{eq:forcefinal} that reproduces equation \eqref{eq:energy} for any hyperbolic Kepler orbit and lead the assumption to a contradiction. It turns out, that it is sufficient to deal with a specific class of polar orbits, namely orbits which go from south to north in such a way that on the incoming branch and during the passage of the perigee $\dot\delta$ is positive and on the outgoing branch the spacecraft passes the pole ($\delta = \frac{\pi}{2}$). In the following, we will restrict our considerations to these polar flybys, for those there is
	\begin{align}
		\label{eq:cosdiff}
		\cos\delta_i - \cos\delta_o & = -2\sin\left(\frac{\delta_i - \delta_o}{2}\right)\sin\left(\frac{\delta_i + \delta_o}{2}\right) = 2\frac{1}{\varepsilon}\cos\delta_p
	\end{align}
with $\delta_p$ being the geocentric declination for the closest-approach location. By that we have applied the general relation $\frac{1}{\varepsilon} = \sin\frac{\Delta}{2}$ between the numerical eccentricity $\varepsilon$ and the deflection angle $\Delta$  of a hyperbola. Remarkably, we see that $\omega_E R_E\cos\delta_p = v_{p}$, where $v_{p}$ defines the tangential rotation velocity of the Earth at the perigee. Hence, the specific energy change depends directly on this velocity at the perigee, what will be discussed later on more explicitly in Section 4.

According to $E = \frac{1}{2}V_\infty^2 = \frac{1}{2}(\varepsilon - 1)\frac{GM}{r_p}$ where $r_p$ is the value of the radius coordinate $r$ at the perigee, equations \eqref{eq:energy} and \eqref{eq:cosdiff} lead to
	\begin{align}
		\label{eq:DE1}
		\Delta E & = 2K\frac{\varepsilon - 1}{\varepsilon}\frac{GM}{r_p}\cos\delta_p\,.
	\end{align}
On the other hand, we can express the energy change by inserting a field of the form \eqref{eq:forcefinal} into equation \eqref{eq:energy} which gives
	\begin{align}
		\label{eq:DE2}
		\Delta E & = KGM\int\limits_{-\infty}^\infty\frac{f(\delta)}{r^2}\dot r\dd t\,.
	\end{align}
Introducing the polar angle $\varphi$ for the parametrization of the hyperbola (see \eqref{eq:rparam}), we can write
	\begin{align}
		\frac{\dot r}{r^2} = -\totald{}{t}\left(\frac{1}{r}\right) = -\totald{}{t}\left(\frac{1 + \varepsilon\cos\varphi}{(1 + \varepsilon)r_p}\right) = \frac{\varepsilon}{1 + \varepsilon}\frac{1}{r_p}\sin\varphi\,\dot\varphi.
	\end{align}
With that the comparision of the right-hand sides of equations \eqref{eq:DE1} and \eqref{eq:DE2} result in
	\begin{align}
	\label{eq:epsilon}
		\frac{\varepsilon^2 - 1}{\varepsilon^2}2\cos\delta_p & = \int\limits_{\varphi_i}^{\varphi_o} f(\delta)\sin\varphi\dd\varphi\,.
	\end{align}
For $\varepsilon \to \infty$, whereby the hyperbolas degenerate to straight lines, we get
	\begin{align}
		2\cos\delta_p & = \int\limits_{-\frac{\pi}{2}}^{\frac{\pi}{2}}f(\delta)\sin\varphi\dd\varphi\,.
	\end{align}
For our special polar flybys this limit yields $\varphi = \delta - \delta_p$ for $-\frac{\pi}{2} \le \varphi \le \frac{\pi}{2} - \delta_p$ and $\varphi = \pi - \delta - \delta_p$ for $\frac{\pi}{2} - \delta_p \le \varphi \le \frac{\pi}{2}$. Consequently, we should divide the integration according to the borders
	\begin{align}
	\label{eq:intsep1}
		2\cos\delta_p & = \int\limits_{-\frac{\pi}{2} + \delta_p}^{\frac{\pi}{2}}f(\delta)\sin(\delta - \delta_p)\dd\delta - \int\limits_{\frac{\pi}{2}}^{\frac{\pi}{2} - \delta_p}f(\delta)\sin(\pi - \delta - \delta_p)\dd\delta\\
	\label{eq:intsep2}
									& = \cos\delta_p\left(\int\limits_{-\frac{\pi}{2} + \delta_p}^{\frac{\pi}{2}} - \int\limits_{\frac{\pi}{2}}^{\frac{\pi}{2} - \delta_p}\right)f(\delta)\sin\delta\dd\delta - \sin\delta_p\int\limits_{-\frac{\pi}{2} + \delta_p}^{\frac{\pi}{2} - \delta_p}f(\delta)\cos\delta\dd\delta\,.
		\end{align}
Since $f$ can assumed to be a symmetric function (see Step 5 above) the part $\delta\in \left[-\frac{\pi}{2} + \delta_p,\frac{\pi}{2} - \delta_p\right]$ of the first integral vanishes. After division by $2\cos\delta_p$, equations \eqref{eq:intsep1} and \eqref{eq:intsep2} reduce to 
\begin{align}
1 & = \int\limits_{\frac{\pi}{2} - \delta_p}^{\frac{\pi}{2}}f(\delta)\sin\delta\dd\delta - \tan\delta_p\int\limits_{0}^{\frac{\pi}{2} - \delta_p}f(\delta)\cos\delta\dd\delta\,.
\end{align}
Furthermore, for $\delta_p \to 0$ the right-hand side of the equation vanishes as well and the whole equation results in the inconsistency $1 = 0$. Hence, we see that the assumption there is a velocity-independent force field cannot be correct.

It may be noted that, because of the antisymmetry of $\Delta E$ as well as $\int\vec V\dd t$ under the transformation $t\rightarrow -t$, terms with velocity dependence of odd-numbered power cannot contribute to the anomalous energy change $\Delta E$. For scaling reasons the radius has the power of $p-4$ when the velocity has the power of $p$. For any $p > 2$ the force field would not vanish for increasing radius. Therefore, other values for $p$ than 0 and 2 are out of the question.

\section{Velocity-dependent solutions}

Since we have seen that there is no velocity-independent force field reproducing equation (1) we will now put up with forces depending on the spacecraft´s velocity. We will see that intuitional transformations of \eqref{eq:caltech} lead to a wide varity of fields which exactly model the flyby anomaly.

An obvious idea to find velocity-dependent fields $\vec F(\vec r,\vec V)$ solving \eqref{eq:energy} is to regard the right-hand side as the difference of the values of an antiderivate of a suitable function, evaluated at the ``in'' and ``out'' values of the integration variable, and ``read off'' the field from the integrand. In the following we can resort to the fact that the specific energy $E$ is a constant value along any undisturbed trajectory, but in order to get a field which depends only on the variables $\vec r$ and $\vec V$ (and in order to avoid obtaining a family of fields, parametrized by $E$), we have to express $E$ by $\vec r$ and $\vec V$ through $E = \frac{1}{2}V^2 - \frac{GM}{r}$.

Formally, this approach gives for the present
	\begin{align}
		\label{eq:intinterpret}
		\Delta E & = -K\int\limits_{-\infty}^\infty\left(V^2 - \frac{2GM}{r}\right)\totald{\gap(\cos\delta)}{t}\dd t\,,
	\end{align}
but there is an ambiguity concerning the meaning of the $\delta$ coordinate. Only their special values $\delta_i$ and $\delta_o$ are well defined by the directions of the asymptotic velocity vectors. However, since the asymptotic directions of the velocity vectors coincide with the asymptotic geocentric positions of the spacecrafts, we have the choice between the ``velocity-space interpretation'' and the ``configuration-space interpretation''. The former interpretation leads via
	\begin{align}
		\totald{}{t}\cos\delta = \totald{}{t}\sqrt{1 - \frac{V_z^2}{V^2}}\,,
	\end{align}
where is $V_z := V\sin\delta$, to quite ``unnatural'' fields with a complicated $\vec V$ dependence and components of $\vec V$ in the denominator, whereas the latter interpretation deals with $\delta$ as a coordinate (a function) in configuration space and gives
	\begin{align}
		\label{eq:nablacos} 
		\Delta E & = -K\int\limits_{-\infty}^\infty\left(V^2 - \frac{2GM}{r}\right)(\nabla\cos\delta)\cdot\vec V\dd t\\
	\label{eq:sin}
		& = -K\int\limits_{-\infty}^\infty\left(\frac{V^2}{r} - \frac{2GM}{r^2}\right)\sin\delta\,\vec e_\delta\cdot\vec V\dd t\,.
	\end{align}
The comparision with \eqref{eq:energy} shows us that
	\begin{align}
		\label{eq:forcefield}
		\vec F(\vec r,\vec V) & = K\left(\frac{V^2}{r} - \frac{2GM}{r^2}\right)\sin\delta\,\vec e_\delta
	\end{align}
is a solution of \eqref{eq:energy}. Therefore, it reproduces exactly the formula \eqref{eq:caltech} in the framework of our perturbational approach.

Let us point out a remarkable property of this field. Considering the idea of interpreting \eqref{eq:caltech} as an evaluation of an integral, it can easily be understood that the flyby anomaly affects only unbounded orbits. Since for each Kepler orbit the energy and thus $V^2 - \frac{2GM}{r}$ is a constant, a spacecraft is effectively subject to a field proportional to $\nabla\cos\delta$ (cf. \eqref{eq:nablacos}), which is therefore a conservative field (in the sense that the corresponding 1-form is not only closed, but also exact). Consequently, the integral \eqref{eq:nablacos} vanishes for any \underline{bounded} (elliptical or circular) Kepler orbit (in this case \eqref{eq:sin} comes from a loop integral, and t, of course, extends only over a finite interval). Concerning the Pioneer and flyby anomalies, this is in accordance with a conjecture often pronounced by Claus Lämmerzahl (see, e.g., \cite{lamm}, p. 97), namely that escape orbits behave qualitatively different from bounded ones, whereby the latter show no anomalous effects. (But note that, of course, our field \eqref{eq:forcefield} gives a vanishing energy change in general only after integration over a complete cycle.)

As easily can be seen, the force field \eqref{eq:forcefield} is discontinuous at $\delta = \pm \frac{\pi}{2}$. Especially the divergence of the velocity-independent part goes to infinity at the axis of Earth. In connection with this undesirable properties let us consider the term $\nabla\frac{\cos\delta}{r} = -\frac{\cos\delta}{r^2}\vec e_r - \frac{\sin\delta}{r^2}\vec e_\delta$. Since this term multiplied with an arbitrary constant gives no contribution to the integral \eqref{eq:energy}, we may add $-2wKGM\nabla\frac{\cos\delta}{r}$ to our solution \eqref{eq:forcefield}, where the "weight" $w$ is an arbitrary real number. The resulting one-parameter family of solutions $\vec F_w$ reads
	\begin{align}
		\vec F_w & = -K\left(2(w-1)\frac{GM}{r^2}\cos\delta\,\vec e_r - \left(\frac{V^2}{r} - 2w\frac{GM}{r^2}\right)\sin\delta\,\vec e_\delta\right)
	\end{align}
where is
	\begin{align}
		\vec F_1 & = \vec F\qquad\text{and}\qquad \vec F_0 = K\left(\frac{V^2}{r}\sin\delta\,\vec e_\delta + \frac{2GM}{r^2}\cos\delta\,\vec e_r\right)\,.
	\end{align}
(Note that for bounded orbits the property of vanishing total energy change mentioned above holds true as well for any $w$.) The special case $w = 0$ corresponds to the "radialization" procedure in Step 5 of Section 2 and gives a field ($\vec F_0$) with a continuous velocity-independent part. The divergence of this part vanishes everywhere.

However, as we have analyzed, the velocity-\underline{dependent} part of the field \eqref{eq:forcefield} cannot be transformed in a similar way. Independent of the "radialization", the problem with a diverging total field energy in the limit $r \rightarrow \infty$ of that part remains unsolved.

\section{Discussion}

We have analyzed the consequences of the assumptions that the approximation formula given by Anderson et al. \cite{anderson} is valid for all possible spacecraft flyby manoeuvers on the Earth and that the flyby anomaly is a ``real'' effect which can be described by a force field belonging to some unknown interaction between the Earth and the spacecrafts. (The interesting question, whether for flybys on other celestial bodies the coefficient $K$ in equation \eqref{eq:K} is to be replaced by twice the $\beta$ factor of the body's spin rotation velocity at its equator, is beyond the scope of the present paper.) By a ``real'' effect we mean that the conclusion from the radio Doppler data to the anomalous velocity changes is correct and that the anomaly is not (totally or to a considerable part) due to unexplained influences on the propagation of the radio signals or to any kind of mismodelling in the domain of conventional celestial mechanics or other applied branches of physics. Since the mass of the spacecraft does not enter equation \eqref{eq:caltech}, it seems that the velocity changes are caused by unknown gravity-like forces, which, in a classical approach, are to be described by a force field.

As we have seen, such a force field inevitably contains velocity-dependent terms in order to be in accordance with \eqref{eq:caltech}. Equation \eqref{eq:caltech}, however, has been published by Anderson et al. \cite{anderson} as an ``empirical prediction formula'' which may have only approximative character. It remains the possibility that for extreme values of the relevant orbital parameters this formula loses its applicability. To us, it seems reasonable to assume that the symmetry properties we utilized in steps 1 to 3 of section 2 are of general validity. Concerning step 4, one may argue that there may appear deviations for very large perigee distances. As long as such deviations are restricted to the region beyond the observed parts of the asymptotic branches of the hyperbolic orbits, this does not effect our no-go result. Let us also note that the no-go result does not depend on the validity of the approximation formula \eqref{eq:caltech} for arbitrarily large excentricities $\varepsilon$: Coming from \eqref{eq:epsilon}, one can complete the proof without performing the limit $\varepsilon \to \infty$, but compare the powers of $\varepsilon$ in \eqref{eq:epsilon}. However, due to the $\varepsilon$ dependence of the borders of \eqref{eq:epsilon} (which diappears in the limit $\varepsilon \to \infty$), this calculation turns out to be much more involved.

In the proof of our no-go result in section 2 we learned that the total energy change $\Delta E$ remains bounded in the limit of infinitely large excentricity $\varepsilon$, see \eqref{eq:DE1}. (Contrary to a first impression, this is not inconsistent with the fact that, according to formula \eqref{eq:caltech}, $\Delta E$ is proportional to the energy $E$ which diverges in the limit $\varepsilon \rightarrow \infty$, since in this limit the cosine-difference factor goes to zero.) However, despite this property there is the problem of the temporally resolved amount of energy transfer in the course of the passage, which has already been studied by Anderson, Campbell and Nieto \cite{anderson2}. In the case of the field \eqref{eq:forcefield}, the energy transfer integrated over the incoming branch of the hyperbola as well as the transfer on the outgoing branch diverges in the limit $\varepsilon \rightarrow \infty$. Only their sum remains bounded. (Instead of calculating the integrals belonging to the two branches, we can consider the situation within the picture of an effective potential proportional to $E\cos\delta$, where $E$ is to be regarded as a constant. In this picture, it becomes clear that in the limit $\varepsilon \rightarrow \infty$ there are infinitely high potential differences to stride through.)

As a byproduct of the relation \eqref{eq:cosdiff} we see that for the special class of orbits, considered in section 2, the right-hand side of formula \eqref{eq:caltech} reduces to $K(\cos\delta_i-\cos\delta_o) = 4\frac{\omega_E R_E}{\varepsilon c}\cos\delta_p$. Hence, the relative velocity change is proportional to the velocity $v_p$ of the Earth´s surface under the perigee. We want to point out that this interesting relation, which may feign that the energy transfer is caused by the movement of the Earth´s surface directly below the spacecraft, cannot, as we have analyzed, be generalized to nonpolar orbits by fitting in a suitable factor taking into consideration the inclination. Even for polar orbits whose direction of the velocity vector swings over one of the poles ($\delta = \pm \frac{\pi}{2}$), this relation does not hold true.

In section 3 we mentioned that there is, in principle, another possibility to ``intrerpret'' the $\delta$ coordinate in the integral \eqref{eq:intinterpret}. For the following reason it is expected that this way leads to fields with vastly different energy transfer properties: Whereas in our ``configuration-space interpretation'' $\delta$ varies over more than half the celestial sphere, starting from $\delta_i$ until it reaches its final value $\delta_o$, in the ``velocity-space interpretation'' the respective arclength is much shorter, namely given by the total deflection angle. Moreover, the "radialization" of the velocity-independent part of the field (end of Sect. 3) affects the energy transfer process. This may be relevant in view of an observation made by Anderson, Campbell and Nieto \cite{anderson2}: The energy transfer process does not consist of monotonic increase (or decrease) of energy with time but is rather more complicated.

Additionally to the energy transfer problem, which is due to the lack or very small number of data points between the asymptotic branches of the hyperbolas, all our efforts to infer the acting forces from the measured anomalous velocity changes suffer from the lack of any information about the force components perpendicular to the direction of motion of the spacecrafts. So we must realize that nowadays there is no unique solution of our problem (unless we make further strong assumptions), and we are restricted to more general statements, especially about the velocity-dependence.

Let us note that, quite generally, any velocity-dependent force requires further forces due to the Galilei invariance (or Lorentz invariance). In our case, at any point in space we can transform the $V^2$ part of the field to zero by a Galilei transformation, whereby the whole force has to be invariant. That implies nontrivial transformation properties of the velocity-independent part of the field. This shows us that both parts may not be viewed as independent fields and that the velocity-independent part has a structure which differs substantially from the common Newtonian gravity field.

In view of this close interlocking of both parts of the field, we should not overrate the divergence properties of the velocity-independent part analyzed in section 3. (After the "radialization", this part can be interpreted as a gravity shielding which is most effective in the Earth´s equatorial plane.)

\section*{Acknowledgment}

W. Hasse wishes to thank Volker Perlick (Lancaster University) for helpful discussions.

\section*{Contact}

\begin{tabular}{ll}
	Emrah Birsin		& birsin@physik.hu-berlin.de\\[.2cm]
	Philipp Hähnel	& haehnel@physik.hu-berlin.de\\[.2cm]
	Wolfgang Hasse	&	astrometrie@gmx.de	
\end{tabular}

\end{document}